\documentclass[aps,pra,showpacs,twocolumn,amsmath,amssymb,superscriptaddress,footinbib]{revtex4}

\usepackage[english]{babel}
\usepackage{latexsym}
\usepackage{graphics,psfrag}
\usepackage{epsfig}
\usepackage[T1]{fontenc}
\usepackage[usenames,dvipsnames]{color}
\usepackage{ulem}

\def\be{\begin{equation}}
\def\ee{\end{equation}}
\def\bea{\begin{eqnarray}}
\def\eea{\end{eqnarray}}
\def\bi{\begin{itemize}}
\def\ei{\end{itemize}}
\def\bin{\begin{enumerate}}
\def\ein{\end{enumerate}}

\begin{document}

\title{Dipolar bosons on an optical lattice ring}

\author{Micha\l{} Maik}

\affiliation{
Instytut Fizyki imienia Mariana Smoluchowskiego, 
Uniwersytet Jagiello\'nski, ulica Reymonta 4, PL-30-059 Krak\'ow, Poland}

\author{Pierfrancesco Buonsante}
\affiliation{Dipartimento di Fisica, Universit\`a degli Studi di Parma, V.le G.P. Usberti n.7/A, 43100 Parma, Italy}
\author{Alessandro Vezzani}
 \affiliation{Centro S3, CNR Istituto di Nanoscienze	, via Campi 213/a, 41100 Modena, Italy}
\affiliation{Dipartimento di Fisica, Universit\`a degli Studi di Parma, V.le G.P. Usberti n.7/A, 43100 Parma, Italy}

\author{Jakub Zakrzewski$^{1,}$} 

\affiliation{
Mark Kac Complex Systems Research Center, 
Uniwersytet Jagiello\'nski, Krak\'ow, Poland}

\date{\today}

\begin{abstract}

We consider an ultra-small system of polarized bosons on an optical lattice with a ring topology interacting via long range dipole-dipole interactions. Dipoles polarized perpendicular to the plane of the ring reveal sharp transitions between different density wave phases. As the strength of the dipolar interactions is varied the behavior of the transitions is first-order like.  For dipoles polarized in the plane of the ring the transitions between possible phases show pronounced sensitivity to the lattice depth. The abundance of possible configurations may be useful for quantum information applications.

\end{abstract}


\maketitle

The behavior of atoms in optical lattice potentials has attracted a lot of attention in the past decade \cite{lewen07}.  The impressive degree of control over atomic samples gives unprecedented chances for studying models coming from condensed matter physics, e.g. the Hubbard model for fermions or the Bose-Hubbard model for bosons \cite{jaksch05}.  This allows an investigation of relatively large systems which are needed for quantum phase transition problems \cite{greiner02} in order to minimize finite size effects.  A subsequent progress reaches the stage of single atom or single site imaging \cite{bakr09}.  Also optical lattices of practically arbitrary configurations can be prepared \cite{lat2,Amico05} opening up new possibilities.

While ultra-cold atom-atom interactions are typically dominated by contact terms, long range dipole-dipole interactions are  becoming of great interest - for recent reviews see \cite{dipoles,dipole2}. Experimental realizations of condensates of Cr \cite{griesmaier05} and very recently of dysprosium \cite{lu11} open up a pathway for novel physics of atomic species exhibiting strong dipolar interactions. On the other hand, owing to tunable Feshbach resonances one is able to  control scattering lengths between particles very precisely \cite{chin10}.  This tuning may help also to enhance the role of dipolar interactions.
 
Quite impressive work has been performed on polarized dipolar gases in one-dimensional lattice systems
(see e.g. \cite{goral02,altman,deng11,1DBoseGas2,1DBoseGas3,1DBoseGas4}). Here by using the density matrix renormalization group (DMRG), quantitative studies  have been carried out efficiently even for relatively large systems. New phases such as density waves (absent for purely on-site interactions as in the Bose-Hubbard model) can then be observed \cite{goral02,altman}. Longer range interactions allow for the presence of the Haldane insulating phase (with a non-trivial so-called string order \cite{altman,deng11}). In these works it is  assumed that the dipoles are polarized perpendicularly to the one-dimensional lattice so the dipolar interaction is repulsive. This way one can avoid possible instability due to ``head-tail'' attraction between dipoles.

The importance of the dipoles' orientation has been nicely visualized in the study of a dipolar condensate placed in a toroidal trap \cite{abad10}. Assuming the dipoles to be oriented in the plane of the ring it becomes apparent that the dipole-dipole interactions become non-uniform along the ring, with repulsive interactions where locally the ring is perpendicular to the polarization axis and attractive where the ring reinforces ``head-tail'' alignment. Using the mean field Gross-Pitaevskii description it has been shown \cite{abad10} that sufficiently strong dipole-dipole interactions split the condensate into two parts. Another study \cite{zoellner11} considered very few bosons or fermions polarized at an arbitrary angle with respect to a small ring by using exact diagonalization. In both these works no additional optical lattice was assumed.

The results presented in this paper combine the two situations. We consider a few bosons on an optical lattice realized with a ring topology and consider dipolar interactions with different polarizations. Interestingly, this leads to qualitatively novel behavior of the model.

Assuming the lattice to be sufficiently deep so that the excited ``bands'' may be neglected \cite{jaksch05} (for limitations on that approximation see e.g. \cite{luehmann11,sowinski11}), the system may be described by an extended Bose-Hubbard model:

\be
\hat{H} = \displaystyle\sum\limits_{i}\left[ -J(\hat{b}_{i} ^{\dagger} \hat{b}_{i+1} + h.c.) + \frac{U}{2}\hat{n}_{i} (\hat{n}_{i}-1)\right] + \displaystyle\sum\limits_{ij} V_{ij} \hat{n}_{i} \hat{n}_{j}.
\label{hamil}
 \ee

\noindent The first part of the Hamiltonian is just the usual one dimensional Bose-Hubbard model, where $\hat{b}_{i}^{\dagger}$ and $\hat{b}_{i}$ are the lattice  creation and annihilation operators and $\hat{n}_{i} = \hat{b}_{i} ^{\dagger} \hat{b}_{i},$ is the corresponding number operator.  The coefficients $J$ and $U$ correspond to the strength of hopping and contact on-site interaction terms.  

The last term in the Hamiltonian describes the dipole-dipole interactions which provide coupling between all the sites. The coefficient $V_{ij}$ depends both on the mutual orientation between two sites and the distance between them:

\be 
V_{ij} = \int\int d\mathbf{r} d\mathbf{r'} w_{i}^{*} (\mathbf{r}) w_{j}^{*} (\mathbf{r'}) \left[\frac{\mu_{0}\mu^{2}} {4\pi} \frac{1-3\cos^{2} \theta}{|\mathbf{r}-\mathbf{r'}|^{3}} \right] w_{j} (\mathbf{r}) w_{i} (\mathbf{r}), 
\ee

\noindent where $w_{i} (\mathbf{r})$ is a Wannier wavefunction localized on site $i$, $\mu_{0}$ is the permeability of free space, and $\mu$ is the magnetic moment of the specific particle
(to be specific we consider magnetic interactions, although a similar analysis holds for electric dipoles). $\theta$ is  the angle between the axis of polarization and the vector connecting two dipoles $\mathbf{r-r'}$ where the dipoles are positioned at $\mathbf{r}$ and $\mathbf{r'}$. Since the interactions are anisotropic, they can be either attractive or repulsive depending on the relative position of the dipoles. It is exactly here where the ring topology of the lattice makes a difference - otherwise, the Hamiltonian, Eq.(\ref{hamil}), is identical to the one-dimensional models considered previously \cite{altman,deng11}. It is also worth stressing that we consider dipole interactions between all sites while large scale DMRG treatments were typically limited to the next-nearest neighbors couplings.

The on-site $i=j$ contribution of the dipole-dipole interaction is assumed  to be incorporated into the $U$ term, while for $i \neq j$ we assume that the lattice is sufficiently deep that the angle $\theta$ and distances may be measured with respect to the centers of the sites (i.e. taken out of the integrals). Defining the lattice constant $a$ as the distance between nearest neighbors $a=|\mathbf{R}_{i}-\mathbf{R}_{i+1}|$ we may construct dimensionless positions of sites along the ring $\tilde {\mathbf R}_i = {\mathbf R}_{i}/a$. Then
\be
V_{ij} =  V \frac{1-3\cos^{2} \theta_{ij}}{|\tilde {\mathbf R}_i-\tilde {\mathbf R}_j|^{3}}
\ee
where $V$ measures the strength of the dipole-dipole interactions and $\theta_{ij}$ corresponds to the angle between the orientation of the dipoles and the vector connecting them at sites $i$ and $j$. 

Consider first the case of dipoles polarized in a direction perpendicular to the plane of the ring. The angle is always $\theta_{ij}=\pi/2$ and the dipoles repel each other and the magnitude of the interaction depends only on the distance between sites via the $1/r^3$ law. In this situation the model becomes quite similar to the one-dimensional linear models with periodic boundary conditions (with quantitative differences coming from different $V_{ij}$ dependence on the distance $i-j$ due to a different geometry). Contrary, however, to typical DMRG studies of large linear systems \cite{altman,deng11} we shall consider 8 bosons sitting on 8 sites (i.e. with unit mean density). This allows us to use standard exact diagonalization routines \cite{exdiag}. 
 
Of course for such a small system we should not speak about quantum phase transitions but merely consider crossovers between different ground state arrangements (note that quite a small dipole system was discussed recently for a triple well potential \cite{lahaye10}). Surprisingly, standard measures (such as a superfluid fraction or a condensate fraction) used to describe large systems are also applicable and useful for our case. At the same time finite size effects become important and interesting. The condensate fraction, $f_c$, is obtained from the largest eigenvalue of the one-body density matrix (OBDM). Observe that for our system $f_c\ge 1/8$. Following Roth and Burnett \cite{roth03} the superfluid fraction, $f_s$, is obtained by including Peierls phase factors in the Hamiltonian to simulate the slow motion in the system (in our case this corresponds to a spinning of the ring).

The Hamiltonian, (\ref{hamil}), depends on three parameters, $J$, $U$, and $V$.  We assume $J=1$ thus expressing 
$U$, $V$, and $H$ in units of $J$.

\begin{figure}
\begin{center}
\includegraphics[width=0.9\columnwidth]{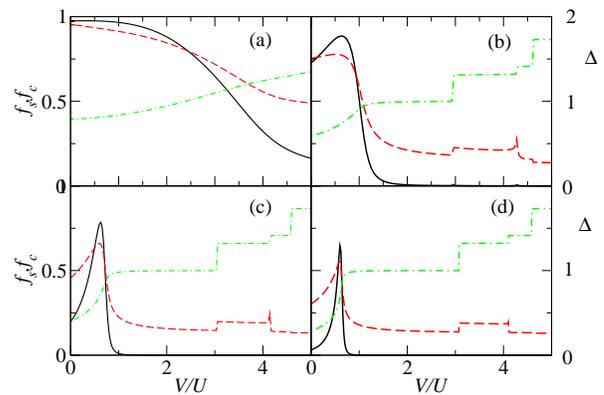}
\end{center}
\caption{(color online) 
Superfluid fraction $f_s$ (solid black lines), condensate fraction $f_c$ (red dashed lines) both represented on left-side vertical axis as a function of $V/U$ for $U=1,4,7,10$ (panels (a) to (d), respectively). Right axis scale corresponds to green dot-dashed lines that represent variance of occupation on different sites. For small $V/U$ we observe an increase of the superfluid (and condensate) fraction with $V$ both in the SF phase (for $U=1$ panel (a) as well as for $U=10$ panel (d)) where the increase of $V$ causes transition from MI to SF phase. Around $V/U=0.5$ a transition to a density wave state $[{2,0}]$ occurs,
more sudden for larger $U$, as manifested by a rapid drop of $f_s$ and $f_c$.
For larger $V/U$ sharp transitions between different density waves occur as
manifested in $f_c$ as well as in variance of occupation $\Delta$. 
}
\label{fig:op1}
\end{figure}

Fig.~\ref{fig:op1} presents $f_s$, $f_c$, as well as the variance of the occupation, $\Delta=\sqrt{\langle n_i^2 \rangle - \langle n_i \rangle^2}$ as a function of the ratio of  $V$ and $U$.
Different panels refer to different values of $U$. In particular, panels (c) and (d) correspond to values of $U$ at which the infinite system at $V=0$ would be deep in the gapped Mott insulating phase. A rough explanation of the behaviour of $f_s$ and $f_c$ at small $V/U$ can be obtained assuming that the hopping amplitude is negligible, $J\approx 0$. At $V=0$  a gap  $\Delta E = U$ separates the ground state energy from the first excited state. In this situation one would expect $f_s = f_c = 0$, and the observed  non-zero values are due to finite-size effects.  A simple calculation shows that for small values of $V/U$ the gap is reduced to $\Delta E = U-V$, which entails an increase in the superfluid and condensate fraction. 

This increase in the superfluid and condensate fraction is not observed in panel (a), corresponding to a value of $U/J$ at which the infinite system would be superfluid in the absence of dipolar interactions. These quantities start off already at their maximum value, and decrease as $V$ is increased. Panel (b) refers to $U/J=4$, which would be slightly inside the Mott insulating state for an infinite lattice at unitary filling and in the absence of dipole interactions. Indeed the behaviour of the finite system under examination is more alike to that observed at larger $U/J$, in panels (c) and (d), although not as sharply peaked.

In panels (b)-(d)  the SF fraction exhibits a sudden decrease with increasing dipole interaction at $V/U\approx 0.5$, and the system enters into a density-wave state characterized by a site occupation number alternating between $0$ and $2$. We denote this situation as $[2,0]$ state. As we discuss in the following, different density-wave states take over at larger values of $V/U$. As can be seen from the plots in panels (b)-(d), the (sharp) crossover points between subsequent density-wave regimes are essentially independent of the value of $U$, since at negligible $J/U$ the only relevant parameter is $V/U$. 

Since all sites are equivalent, the system is invariant under rotation by $2\pi/M$ where $M=8$ is the number of sites. The ground state shares this symmetry. There are two almost degenerate lowest lying  states corresponding to even and odd combinations of $[2,0]$ and $[0,2]$ states satisfying $\langle n_i \rangle =1 $ at every site. To identify the density wave state, we construct the density matrix of the ground state and trace out all sites except one. This yields the probability distribution $P(n)$ of bosons at
a single site (see Table~\ref{tab}). The second column (for which the $V/U$ value corresponds to a $[2,0]$ density wave phase) shows that $n=0$ and $n=2$ are occupied with the same probability being almost 1/2 confirming
the alternating empty and doubly occupied sites, the remaining $n$ values have non vanishing occupation due to finite tunneling. Observe that just below  $V/U=3$ (see fig.~\ref{fig:op1}b) a jump appears in $f_c$ and the variance $\Delta$ - a transition occurs to the density wave state in which two sites are occupied by 3 bosons and one by 2 bosons, as it is evident from $P(n)$ distribution in the third column of  Table~\ref{tab}. The next jump in $\Delta$ indicates a transition to one site occupied by 4 bosons and two by 2 bosons while the last jump creates two sites with 4 bosons with the rest remaining empty. Eventually for much higher $V/U$ (not shown) one observes that all bosons accumulate at a single site, leaving 7 sites empty. This is because accumulation of particles at one site costs little energy (proportional to $U$ - a vanishingly small parameter in this limit) while occupation of any two sides gives a nonzero term proportional to $V$. This behaviour is in a sharp contrast with a cristal-like structure obtained for a ring topology without a lattice \cite{zoellner11} where the dipole interaction enhances repulsion due to a contact term thus small distance between bosons are avoided. Recall, however, that we assume that the dipole-dipole interaction between particles at a single site is incorporated in $U$ term already, so there is no additional energy costs for particles to accumulate at a single site.
\begin{table}[h]
\centering
\begin{tabular}{|l|c|c|c|c|}
\hline
 n & $V/U=2.5$ & $V/U=3$ &  $V/U=4.5$ & $V/U=5$\\
\hline
 0 & 0.4950 & 0.6139  & 0.6225 & 0.7491 \\
 1 & 0.0124 & 0.0181  & 0.0057 & 8.6 $10^{-4}$ \\
 2 & 0.4902 & 0.1229  & 0.2456 & 3.6 $10^{-6}$ \\
 3 & 0.0024 & 0.2453  & 0.0018 & 8.6 $10^{-4}$\\
 4 & 2.3 $10^{-6}$  & 0.0049 & 0.1244 & 0.2491 \\
\hline
\end{tabular}
\caption{Probability distributions at a given site for different ratio of dipolar to on-site coupling,
$V/U$ for $U=4$ revealing different density wave arrangements.}
\label{tab}
\end{table}

Naturally, the types of density waves observed are strongly system size dependent. For 9 bosons on 9 sites
we observed a state with one site occupied by 3 bosons and three sites by 2 bosons, then a crossover into a repetition of $[3,0,0]$ sequence and for large $V/U$ one site with 5, another with 4 atoms and the remaining sites empty.
It should be noted that some density waves taking over the ground state of the system appear because truly long-range interactions are considered. If, as chosen 
in \cite{goral02,altman,deng11}, the range of the interaction is truncated, some density-wave configurations never crop out. For instance, if no interactions are included beyond next-nearest-neighbors, the configuration  where all bosons accumulate at a single site cannot appear, no matter how strong the interaction is.
 Importantly observe that crossovers between different density waves resemble phase transitions of the first kind (the jump in $\Delta$ and also in $f_c$).

\begin{figure}[h]
\begin{center}
\includegraphics[width=0.95\columnwidth]{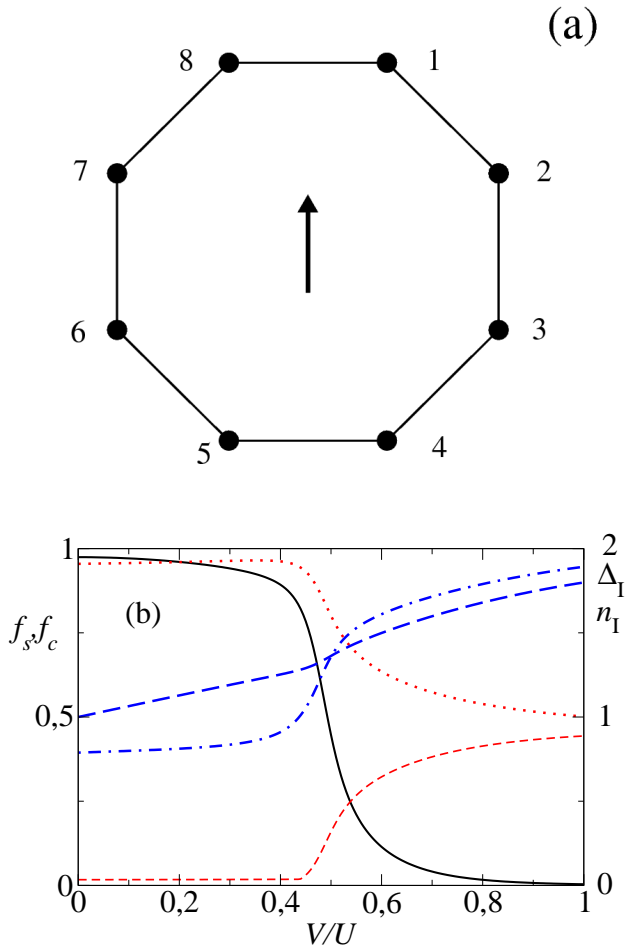}\\
\includegraphics[width=0.95\columnwidth]{U1in.eps}
\end{center}
\caption{(color online) 
A ring shaped lattice (a), the arrow indicates polarization of dipoles. Due to symmetry sites 2,3,6,7 are equivalent (type-I), similarly 1,4,5,8 (type-II). Panel (b) shows results for a shallow lattice, $U/J=1$. The solid black line shows a SF fraction with a  drop around $V/U=0.5$. This drop is accompanied by similarly rapid increase of the number variance $\Delta_I$
(blue dash-dotted line, right axis scale) as well as changes of condensate fraction, $f_c$ (dashed red line) and the second eigenvalue of OBDM, $g_c$ (dotted red line). On the other hand the mean occupation $n_I$ of type-I sites (blue dashed line, right axis scale) increases slowly and smoothly in the whole studied range of $V/U$. For the discussion see text.
}
\label{fig:in1}
\end{figure}

Consider now the the orientation of dipoles {\it in the plane} of the ring. In this case interactions between two sites are not only dependent on the distances separating them but are also influenced by their relative positions on the lattice. Due to the small size of our system (we again consider 8 bosons on 8 sites) different orientations of the dipole polarization with respect to the octagon are possible, the results depend quantitatively on this orientation. Further on we choose the most symmetric case when the
polarization axis coincides with two edges of the octagon (compare Fig.~\ref{fig:in1}a). Then, from symmetry considerations we have
two types of sites: 4 located at the end of the edges that are parallel to the polarization axis
(type-I) and 4 located  at the end of the edges perpendicular to that axis (type-II). Due to the symmetry, sites of a given type are equivalent.
Recall that the dipole-dipole interaction terms in the last sum in Eq. \eqref{hamil} can be either repulsive or attractive. As we will discuss in the following, the physics of the system is dominated by the strong attractive interaction between dipoles occupying neighbouring type-I sites.

\begin{table}
\centering
\begin{tabular}{|l|c|c|c|c|}
\hline
& \multicolumn{2}{|c|}{ Type-I  } &
\multicolumn{2}{c|}{ Type-II  } \\
 \hline
 n & $V/U=0.4$ & $V/U=0.6$ &  $V/U=0.4$ & $V/U=0.6$\\
\hline
 0 & 0.2098 & 0.4839  & 0.4146 & 0.6072 \\
 1 & 0.4260 & 0.0397  & 0.4391 & 0.2955 \\
 2 & 0.2771 & 0.1279  & 0.1313 & 0.0867 \\
 3 & 0.0766 & 0.2132  & 0.0144 & 0.0100\\
 4 & 0.0099 & 0.1140 & 0.0006 & 0.0005 \\
\hline
\end{tabular}
\caption{Probability distributions at sites of type I and II on either side of the $V/U=0.5$ transition
for $U/J=1$, revealing ground state character change.}
\label{tab2}
\end{table}

Fig.~\ref{fig:in1}b shows system properties for a shallow lattice $U/J=1$. 
The increase of $V/U$ has the effect of transferring particles from sites of type II to sites of type I. 
For $V/U\approx 0.5$ a change of the system's properties appears. The SF fraction drops from values close to unity to almost zero; the variance of the particle number in type-I sites, $\Delta_I$, exhibits a sudden and significant increase; similarly a drop in the CF, $f_c$, is observed, while a second eigenvalue of the OBDM becomes large. This suggests that, as $V/U$ exceeds $0.5$ the system essentially consists of the superposition of two coherent halves, each localized  at a pair of neighbouring sites of type I. This superposition of ``macroscopic''states is typical of a ``double well'' system where an effective attractive interaction is dominant \cite{spekkens99,Jack05}, and gives rise to the sudden increase in the number fluctuations observed in Fig.~\ref{fig:in1}b.

This phenomenology is reminiscent of the observations of Ref.~\cite{abad10}, where a situation similar to ours, but without the optical lattice, is analyzed adopting a Gross-Pitaevskii mean-field description. Similar to our case, an increase in the dipolar interaction causes at first a transfer of population around  ``hour 3 and hour 9'', i.e. regions where dipoles arranged head to tail attract each other. As the interaction is further increased a symmetry breaking -- typical of attractive nonlinear systems -- occurs, and the entire boson population localizes either around ``hour 3'' or ``hour 9''. This symmetry broken situation is the mean-field counterpart of the macroscopic superposition occurring at the quantum level.
The above description is confirmed by the inspection of the one site occupation probability distribution, $P(n)$. Table II presents $P(n)$ for values of $V/U$ slightly below and slightly above  $0.5$. For  type-II sites $P(n)$ changes smoothly.
On the other hand $P(n)$ for type-I reveals that the for $V/U>0.5$ the state becomes a superposition of the vacuum with a state centered around larger $n$. 

For larger values of the dipolar interaction ($V/U>30$) the state of the system essentially consists of the superposition of two Fock states, $\frac{1}{\sqrt{2}}|0, 4, 4, 0, 0, 0 ,0 ,0\rangle + \frac{1}{\sqrt{2}}|0, 0, 0, 0, 0, 4 ,4 ,0\rangle$, whose condensate fraction is significantly smaller than $1/2$.

\begin{figure}[h!]
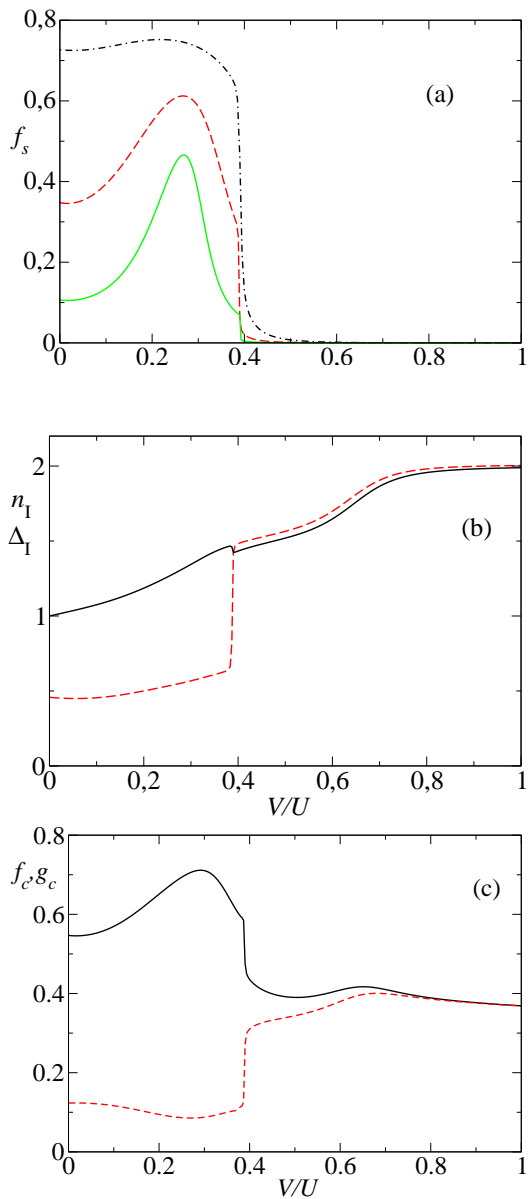

\begin{center}
\begin{tabular}{c}
\includegraphics[width=0.8\columnwidth]{Uintera.eps} \\  \\
\includegraphics[width=0.8\columnwidth]{Uinterb.eps} \\
\includegraphics[width=0.8\columnwidth]{Uinterc.eps} 
\end{tabular}
\end{center}
\caption{(color online) 
Intermediate $U/J$ behavior. SF fraction is shown in panel (a) for $U/J=4$ (black dotted line), $U/J=6$ (red dashed line) and 
$U/J=8$ (solid green line). Panel (b) presents the mean occupation and the variance of type-I states for $U/J=6$ while
panel (c) shows the $V/U$ dependence (again for $U/J=6$) of the condensate fraction, $f_c$, i.e. the largest eigenvalue of the OBDM, and the second largest eigenvalue, denoted by $g_c$. The sudden drop of SF fraction occurring at $V/U=0.4$ in panel (a)  is accompanied by a sharp increase of the occupation variance in type-I sites $\Delta_I$ [red dashed line in (b)]. The occupation $n_I$ (black line) shows there a small kink too. At the same $V/U=0.4$ the sharp drop in $f_b$ is observed, while $g_b$ shows an increase.
}\vskip 0.2truecm
\label{fig:in2}
\end{figure}

\begin{figure}
\begin{center}
\null\vskip .4truecm
\includegraphics[width=0.9\columnwidth]{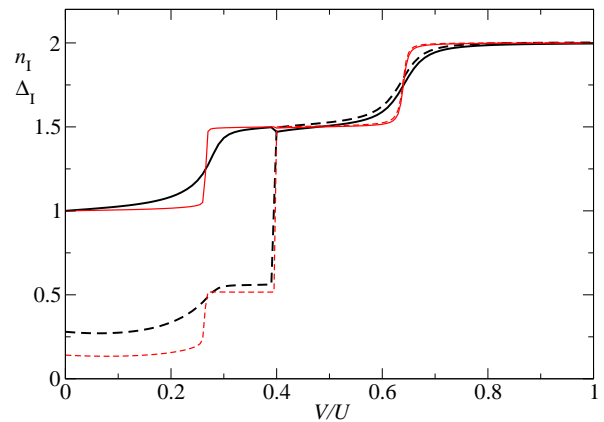}
\end{center}
\caption{(color online) Occupation of type-I sites, $n_I$ for $U/J=10$ (solid black line) and $U/J=20$ (solid red line).
The corresponding variance $\Delta_I$ is plotted for $U/J=10$ (black dashed line) and $U/J=20$ (red dashed line).  For a discussion see text.
}
\label{fig:in3}
\end{figure}

The ``double-well'' behavior becomes even more spectacular for intermediate values of $U/J$ as demonstrated in Fig.~\ref{fig:in2}.
A sharp crossover is observed around $V/U\approx 0.4$ for all the quantities plotted in panels (a)-(c). Like in Fig.~\ref{fig:in1} this separates the situation where a single coherent state is present from the situation where the system consists of the superposition of two coherent halves. The sharpness of the crossover  may be linked to an extremely narrow avoided level crossing involving the ground-state of the system. Further increase of $V/U$ smoothly depopulates type-II sites while the mean population of type-I sites becomes 2. At even larger $V/U$ the ground-state eventually turns into the superposition of the two Fock states discussed above (as manifested by the near degeneracy of the two largest eigenvalues of the OBDM, i.e. $f_c\approx g_c$ in Fig.~\ref{fig:in2}). We observe also that, at small values of $V/U$, the population transfer from type-II to type-I sites is accompanied by an increase in the superfluid and condensate fractions, as it is evident from panels (a) and (c). This effect has the same origin as explained when discussing Fig. \ref{fig:op1}. At the transition point, apart from sharp changes in superfluid and condensate fractions, we observe numerically also a small kink in the
occupation of type-I sites (with, of course, a similar, reversed kink in occupation of type-II states) pointing to the redistribution of densities between type I and type two sites due to a competition of tunneling and interactions. This effect is no longer present for stronger interactions (compare Fig.~\ref{fig:in3}).

For still stronger on-site repulsion $U/J>10$, Fig.~\ref{fig:in3}, the system remains in an insulator-like state with vanishing SF fraction for all $V/U$ values. In this regime the state of the system  is controlled by the competition between the on-site repulsive and long-range dipolar interaction, and essentially consists of the  superposition of a few Fock states. With reference to the $U/J=20$ case in Fig.~\ref{fig:in3}, the first plateau in the plotted quantities corresponds to the usual Mott state, $|1,1,1,1,1,1,1,1\rangle$. Around $V/U=0.3$ this is replaced by $\frac{1}{\sqrt 2}|0,1,1,0,1,2,2,1\rangle+\frac{1}{\sqrt 2}|1,2,2,1,0,1,1,0\rangle$. Next, around $V/U=0.4$ the state $\frac{1}{\sqrt 2}|0,0,0,0,1,3,3,1\rangle+\frac{1}{\sqrt 2}|1,3,3,1,0,0,0,0\rangle$ takes over, which is then replaced by $\frac{1}{\sqrt 2}|0,0,0,0,0,4,4,0\rangle+\frac{1}{\sqrt 2}|0,4,4,0,0,0,0,0\rangle$ around $V/U=0.65$.
We observe that the long-range character of the dipolar interaction is not essential for the above-described behaviour. It is the anisotropic character of dipolar interactions with  attractive interactions between bosons in type-I sites which is responsible for the observed behaviour. Therefore the same qualitative feature would be observed also when a truncation in the range of interactions is included.

To summarize we have shown that a simple model of a few bosons on a ring-shaped optical lattice can show very rich phenomena. At low $V/U$, dipole-dipole interactions tend to enhance the SF character. For dipoles polarized perpendicularly to the ring plane various density wave states have been observed with sharp transitions between them as $V/U$ was varied. For dipoles polarized in the plane of the ring the separation into two
disjoint parts has been observed (as for the toroidal trap without the lattice \cite{abad10}). The presence of the optical lattices creates additional sharp transitions due to rearrangement of dipoles in the parts of the ring where they attract each other. As $J$ may be varied changing the depth of the lattice and the contact interactions strength may be modified using Feshbach resonances, the arrangements discussed should be accessible experimentally.

This opens up a possibility for constructing quantum computer schemes in analogy to that realized for trapped ions \cite{QuantComp1, QuantComp2}. The latter, due to a strong Coulomb repulsion, are arranged with one ion per trap. The system studied
may allow for controlled interactions between atoms by moving them between the sites in a controlled manner.

M.M. and J.Z. gratefully acknowledge enlightening discussions with K. Sacha at early stages of this work.
This work is supported by the International PhD Projects Programme of the Foundation for Polish Science within the European Regional Development Fund of the European Union, agreement no. MPD/2009/6. Support within Polish Government scientific funds for 2009-2012
as a research project is also acknowledged (J.Z.).


\begin{thebibliography}{16}
\bibitem{lewen07} M. Lewenstein et al., Adv. Phys. \textbf{56}, 243 (2007).
\bibitem{jaksch05} D. Jaksch and P. Zoller, Ann. Phys. \textbf{315}, 52 (2005). 
\bibitem{greiner02} M. Greiner, Nature {\bf 415}, 39 (2002).
\bibitem{bakr09} W. S. Bakr et al, Nature {\bf 462}, 74 (2009).
\bibitem{lat2} K. Henderson, C. Ryu, C MacCormick and M. G. Boshier, \textsl{New J. Phys} \textbf{11}, 043030 (2009)
\bibitem{Amico05}L. Amico, A. Osterloh and  F. Cataliotti, Phys. Rev. Lett 95, 063201 (2005)
\bibitem{dipoles} T. Lahaye, C. Menotti, L. Santos, M. Lewenstein and T. Pfau, Rep. Prog. Phys. \textbf{72}, 126401 (2009).
\bibitem{dipole2}C. Trefzger, C. Menotti, B. Capogrosso-Sansone, and M. Lewenstein, arXiv:1103.3145. 

\bibitem{griesmaier05} A. Griesmaier et al., Phys. Rev. Lett. \textbf{94}, 160401 (2005).
\bibitem{lu11} M. Lu, N. Q. Burdick, S. Youn, and B. L. Lev, arXiv:1108.5993

\bibitem{chin10} C. Chin, R. Grimm, P. Julienne, and E. Tiesinga, Rev. Mod. Phys. {\bf 82}, 1225 (2010).
\bibitem{goral02} K. G\'oral, L. Santos, and M. Lewenstein,  Phys. Rev. Lett. \textbf{88}, 170406 (2002).
\bibitem{altman} E. G. Dalla Torre, E. Berg and E. Altman, Phys. Rev. Lett. \textbf{97}, 260401 (2006).
\bibitem{deng11} X. Deng and L. Santos, Phys. Rev. B{\bf 84}, 085138 (2011).
\bibitem{1DBoseGas2} S. M\"uller, et al., arXiv:1105.5015
\bibitem{1DBoseGas3} R. M. Wilson and J. L. Bohn,   \textsl{Phys. Rev. A} \textbf{83}, 023623 (2011)
\bibitem{1DBoseGas4} T. Roscilde and M. Boninsegni, New Journal of Physics, \textbf{12}, 033032 (2010). 
\bibitem{abad10} M. Abad et al., Phys. Rev. A \textbf{81}, 043619 (2010).
\bibitem{zoellner11} S. Z\"ollner, G. M. Brunn, C. J. Pethick and S. M. Reimann, Phys. Rev. Lett. \textbf{107}, 035301 (2011).
\bibitem{luehmann11} D.-S. L\"uhmann, O. J\"urgensen, and K. Sengstock, arXiv:1108.3013.
\bibitem{sowinski11} T. Sowi\'mski et al., arXiv:1109.4782. 
\bibitem{exdiag} J. M. Zhang and R. X. Dong, \textsl{Eur. J. Phys.} \textbf{31}, 591 (2010)
\bibitem{lahaye10} T. Lahaye, T. Pfau, and L. Santos, Phys. Rev. Lett. {\bf 104}, 170404 (2010) 
\bibitem{roth03} R. Roth and K. Burnett, \textsl{Phys. Rev. A} \textbf{68}, 023604 (2003)
\bibitem{spekkens99} R.W. Spekkens and J.E. Sipe, Phys. Rev. A 59, 3868 (1999)
\bibitem{Jack05} M. W. Jack and M. Yamashita, \textsl{Phys. Rev. A} \textbf{71}, 023610 (2005)



\bibitem{QuantComp1} H. Haeffner, C.F. Roos, and R. Blatt, Phys. Rep. \textbf{469}, 155-203 (2008). 

\bibitem{QuantComp2}A. Friedenauer, H. Schmitz, J. T. Glueckert, D. Porras and  T. Schaetz, Nature Physics \textbf{4}, 757-761 (2008) 

\end{thebibliography}
\end{document}